# Clarification on Einstein's model for heat capacity of a solid

Matt Beekman*, Jonathan Fernsler, and Thomas D. Gutierrez

*Department of Physics, California Polytechnic State University, San Luis Obispo, CA 93402*

**Abstract**

Seemingly different interpretations or descriptions of Einstein's model for the heat capacity of a solid can be found in textbooks and the literature. The purpose of this note is to clarify the equivalence of the different descriptions, which all lead to the same Einstein expression for the heat capacity of a solid of $N$ atoms.

*Email: mbeekman@calpoly.edu

Einstein's model for the heat capacity of a solid [1] is historically important because it was one of the first successes of quantum theory, not just applied to solids, but also more generally, illustrating the need for energy quantization to explain experimental results [2]. It is pedagogically important because it serves as a useful and relatively accessible application of key concepts from statistical physics to the solid-state, while simultaneously providing a nice example of the historical progression of physics. Thus, it is commonly taught at one or more points in the standard undergraduate physics curriculum [3-7]. Einstein's model also still finds application in contemporary condensed matter research, where it has been used to describe nearly dispersionless vibrational modes in solids, such as optical phonons [8], the so-called low-energy "rattling" modes of weakly bonded guest atoms in cage-like crystals [9-11], and atomic vibrations in glasses [12]. The purpose of this brief note is to clarify the equivalence of seemingly different descriptions of Einstein's model that are found in textbooks and the literature, which can contribute to confusion for students learning about the model for the first time.

Although Einstein's original paper was published several years prior to the development of what is considered modern quantum mechanics, Einstein's model for heat capacity is most commonly introduced pedagogically as an application of the quantum harmonic oscillator (QHO). Many descriptions and pictures found in textbooks [3, 7, 13, 14] as well as papers in the literature [15, 16] either directly state or tacitly imply that Einstein's model is equivalent to treating each of the $N$ atoms in the solid as an independent 3-dimensional (3D) QHO, such that all atoms have the same classical angular frequency $\omega_E$, or that the $N$ atoms behave as $3N$ 1-dimensional QHOs. For a 1D QHO, the allowed energies are $\varepsilon_{1D} = \left(\frac{1}{2}+n\right)\hbar\omega_E$, and thus the partition function for the 1D oscillator is

$$Z_{1D} = \sum_{n=0}^{\infty} \exp\left[-\frac{\left(n+\frac{1}{2}\right)\hbar\omega_E}{k_B T}\right] = \frac{e^{-\hbar\omega_E/2k_B T}}{1 - e^{-\hbar\omega_E/k_B T}}, \quad (1)$$

where $k_B$ is Boltzmann's constant and $T$ is the temperature. The average thermal energy $\bar{\varepsilon}_{1D}$ per 1D oscillator is then

$$\bar{\varepsilon}_{1D} = k_B T^2 \frac{\partial (\ln Z_{1D})}{\partial T} = \frac{1}{2}\hbar\omega_E + \frac{\hbar\omega_E}{e^{\hbar\omega_E/k_B T} - 1} . \qquad (2)$$

A common progression in reasoning is that the $N$ atoms in the solid behave as $3N$ 1-dimensional (1D) simple harmonic oscillators [3, 5], and the heat capacity $C$ (at constant volume) for $N$ 3D SHOs is simply equal to $3N$ times the heat capacity for the 1D oscillator,

$$C = 3N \frac{d\bar{\varepsilon}_{1D}}{dT} = 3Nk_B \left(\frac{\hbar\omega_E}{k_B T}\right)^2 \frac{e^{\hbar\omega_E/k_B T}}{(e^{\hbar\omega_E/k_B T} - 1)^2} = 3Nk_B \left(\frac{\Theta_E}{T}\right)^2 \frac{e^{\Theta_E/T}}{(e^{\Theta_E/T} - 1)^2} , \qquad (3)$$

where $\Theta_E$ is the characteristic temperature known as the Einstein temperature, such that $\hbar\omega_E = k_B \Theta_E$. Eq. 3 is Einstein's model for the heat capacity of a solid.

Treatments in most textbooks lack the motivation for the exact equivalence of $N$ 3D QHOs and $3N$ 1D QHOs, and the reader is usually expected to accept this result in good faith, or perhaps it is presumed to be obvious [3, 6, 7, 13, 14]. However, the relation $\bar{\varepsilon}_{3D} = 3\bar{\varepsilon}_{1D}$ is a bit more subtle than it appears and may not be obvious to students, in part because a 3D QHO has a different energy spectrum than the 1D QHO, i.e., in the 3D case all levels except the ground state are degenerate. For an isotropic 3D QHO, the time-independent Schrödinger equation is separable and the energy for a single 3D oscillator becomes [17]

$$\varepsilon_{3D} = \left(\frac{3}{2} + n_x + n_y + n_z\right)\hbar\omega_E = \left(\frac{3}{2} + n'\right)\hbar\omega_E , \qquad (4)$$

where $n_x$, $n_y$, and $n_z$ are independent integers, and $n' = n_x + n_y + n_z$. To see that $\bar{\varepsilon}_{3D} = 3\bar{\varepsilon}_{1D}$, consider the partition function for the 3D case,

$$Z_{3D} = \sum_{n_x=0}^{\infty} \sum_{n_y=0}^{\infty} \sum_{n_z=0}^{\infty} \exp\left[-\left(n_x + n_y + n_z + \frac{3}{2}\right)\frac{\hbar\omega_E}{k_B T}\right]$$

$$= \sum_{n_x=0}^{\infty} \sum_{n_y=0}^{\infty} \sum_{n_z=0}^{\infty} \exp\left[-\left(n_x + \frac{1}{2}\right)\frac{\hbar\omega_E}{k_B T}\right] \exp\left[-\left(n_y + \frac{1}{2}\right)\frac{\hbar\omega_E}{k_B T}\right] \exp\left[-\left(n_z + \frac{1}{2}\right)\frac{\hbar\omega_E}{k_B T}\right]$$

$$= \sum_{n_x=0}^{\infty} \exp\left[-\left(n_x + \frac{1}{2}\right)\frac{\hbar\omega_E}{k_B T}\right] \sum_{n_y=0}^{\infty} \exp\left[-\left(n_y + \frac{1}{2}\right)\frac{\hbar\omega_E}{k_B T}\right] \sum_{n_z=0}^{\infty} \exp\left[-\left(n_z + \frac{1}{2}\right)\frac{\hbar\omega_E}{k_B T}\right]$$

$$= (Z_{1D})^3. \tag{5}$$

Alternatively, to make the effects of degeneracy more explicit, the partition function for the 3D QHO can also be written neatly as

$$Z_{3D} = \sum_{n'=0}^{\infty} \frac{(n'+2)(n'+1)}{2} \exp\left[-\left(n' + \frac{3}{2}\right)\frac{\hbar\omega_E}{k_B T}\right], \tag{6}$$

which is also equivalent to $(Z_{1D})^3$. For a given value of $n'$, there will be a degeneracy of $(n'+2)(n'+1)/2$. This weighting indeed gives exactly the number of times each distinct term appears in the triple sum in Eq. 5. This can be established combinatorically [17]. It can also be shown that the symmetry group of the isotropic 3D quantum harmonic oscillator is SU(3) [18] and the degeneracy is given by the dimension of the $D(n', 0)$ representation [19], matching Eq. 6.

From Eq. 5, it follows that [7]

$$\bar{\varepsilon}_{3D} = k_B T^2 \frac{\partial(\ln Z_{3D})}{\partial T} = 3k_B T^2 \frac{\partial(\ln Z_{1D})}{\partial T} = 3\bar{\varepsilon}_{1D}, \tag{7}$$

and for $N$ 3D QHO oscillators the heat capacity is given by Eq. 3. Students may not question the deceptively simple assertion of equivalence of $N$ 3D oscillators and $3N$ 1D oscillators, but that does not necessarily mean it is obvious to them nor that they understand the rationale. Indeed, while the equivalence of $N$ 3D systems to $3N$ 1D systems that leads to Eq. 7 will be true whenever the energy for the 3D case can be written as a sum of three independent but equivalent energies corresponding to independent degrees of freedom (c.f. the particle in a cubic 3D box, also a standard undergraduate problem), it is not in general true for other systems. Simple counterexamples include the anisotropic 3D QHO, and a particle in a rectangular box with unequal edge lengths, $L_x \neq L_y \neq L_z$.

We conclude by noting yet another mathematically equivalent but perhaps more physical satisfying interpretation [5, 10, 20], namely that the approximation in Einstein's model is equivalent to assuming the $3N$ *normal modes*, i.e. collective oscillations of the lattice, rather than the individual *atoms*, are assumed to all have the same oscillation frequency. This is to say the normal modes have no dispersion and the density of states in frequency is a delta function at $\omega_E$, $g(\omega) = 3N\delta(\omega - \omega_E)$. This ensures that $\int_0^\infty g(\omega)d\omega = 3N$, as required for a solid containing $N$ atoms that can each oscillate in three dimensions. Upon quantization of the energy for each normal mode [21], i.e. effectively treating the energy spectrum of each of the $3N$ collective normal modes (rather than the $N$ individual atoms) as equivalent to that of a 1D quantum harmonic oscillator with the same frequency, and application of Bose-Einstein statistics, this approximation also yields the expression in Eq. 3 [20]. This interpretation is physically more satisfying because, unlike the description in the paragraphs above, it does not suppose that the atoms, bonded in the solid, undergo uncoupled oscillations, and is perhaps more in line with the spirit of Einstein's original paper [1].

We believe the above discussion can also be nicely framed as a homework problem for an upper division solid state or statistical physics course. An example of such a problem is provided in Appendix A.

**Appendix A: Suggested Homework Problem**

The Einstein model of lattice dynamics treats a solid as $N$ 3D independent quantum harmonic oscillators. Because the degrees of freedom for each oscillator are independent, this naively suggests that $N$ 3D independent oscillators should have a specific heat equivalent to $3N$ 1D independent quantum harmonic oscillators, which you will show to be true in this problem.

(i) Starting from the known energy of an isotropic 3D quantum harmonic oscillator with natural frequency $\omega_E$, quantum numbers $n_x$, $n_y$, and $n_z$, at temperature $T$, derive a form of the partition function with a single sum over the single index $n' = n_x + n_y + n_z$. When setting up the partition function, be sure to account for the degeneracy of the states when weighting each term in the sum. *Hint*: The time-independent Schödinger equation is separable in $x$, $y$, and $z$. What is the degeneracy of the states as a function of $n'$?

(ii) Show that, when the degeneracy is accounted for, your partition function for the 3D oscillator from part (i) factorizes into a product of three independent partition functions for 1D quantum harmonic oscillators.

(iii) From the partition function in part (ii), derive a form of the average energy and the specific heat for $N$ such 3D oscillators. Show that this matches the expected form of the Einstein model. *Hint*: Recall the relations $\bar{\varepsilon} = k_B T^2 \frac{\partial (\ln Z)}{\partial T}$, $C_V = (d\bar{\varepsilon}/dT)_V$ and note that $\ln Z^a = a \ln Z$.

**References**

[1] Einstein, Ann. Phys. (Leipzig) **22**, 800 (1907).

[2] H. F. Weber, Ann. d. Phys. Chem. **154**, 367 (1875); *ibid*. **154**, 553 (1875).

[3] F. Mandl, *Statistical Physics*, 2nd Ed. (Wiley, Chirchester, 1988).

[4] C. Kittel and H. Kroemer, *Thermal Physics*, 2nd Ed. (Freeman, New York, 1980).

[5] J. R. Hook and H. E. Hall, *Solid State Physics*, 2nd Ed. (Wiley, Chirchester, 1991).

[6] P. Hofmann, *Solid State Physics: An Introduction*, 2nd Ed. (Wiley-VCH, Weinheim, 2015).

[7] S. H. Simon, *The Oxford Solid State Basics* (Oxford University Press, Oxford, 2013).

[8] N. W. Ashcroft and N. D. Mermin, *Solid State Physics* (Saunders College Publishing, Fort Worth, 1970).

[9] R. P. Hermann, R. Jin, W. Schweika, F. Grandjean, D. Mandrus, B. C. Sales, and G. J. Long, Phys. Rev. Lett. **90**, 135505 (2003).

[10] R. P. Hermann, F. Grandjean, and G. J. Long, Am. J. Phys. **73**, 110 (2005).

[11] M. Beekman, W. Schnelle, H. Borrmann, M. Baitinger, Yu. Grin, and G. S. Nolas, Phys. Rev. Lett. **104**, 018301 (2010).

[12] D. G. Cahill, S. K. Watson, and R. O. Pohl, Phys. Rev. B **46**, 6131 (1992).

[13] M. A. White, *Physical Properties of Materials*, 2nd Ed. (CRC Press, Boca Raton, 2012).

[14] D. L. Goodstein, *States of Matter* (Dover, Mineola, 1985).

[15] E. Lagendijk, Am. J. Phys. **68**, 961 (2000).

[16] R. O Pohl, Am. J. Phys. **55**, 240 (1987).

[17] M. Ligare, Am. J. Phys. **66**, 185 (1998).

[18] D. M. Fradkin, Am. J. Phys. **33**, 207 (1965).

[19] B. C. Hall, *Lie Groups, Lie Algebra, and Representations: An Elementary Introduction*, 2nd Ed. (Springer, Berlin, 2015).

[20] J. M. Ramsey and E. A. Vogler, Am. J. Phys. **45**, 583 (1977).

[21] S. C. Johnson and T. D. Guitierrez, *Am. J. Phys.* **70**, 227 (2002).